\documentclass[aps,prl,10pt,
nofootinbib,twocolumn,preprintnumbers]{revtex4-1}
\usepackage{amsmath,amssymb,graphicx}
\usepackage{dsfont}
\usepackage{xcolor}
\definecolor{Mahogany}{rgb}{0.62,0.24,0.15}
\definecolor{DarkRed}{rgb}{0.6,0,0}
\definecolor{DarkGreen}{rgb}{0,0.6,0}
\definecolor{DarkBlue}{rgb}{0,0,0.6}
\definecolor{DarkOrange}{rgb}{0.9,0.5,0.0}
\definecolor{gray}{RGB}{128,128,128}
\usepackage[colorlinks=true,linktocpage=true,linkcolor=DarkBlue,citecolor=DarkGreen,urlcolor=DarkGreen]{hyperref}
\usepackage{slashed}
\usepackage{verbatim}
\usepackage{enumitem}
\usepackage{mathtools}
\usepackage{calc}
\usepackage{float}     
\usepackage{caption}
\usepackage{soul}

\usepackage{upgreek}
\usepackage{bbm} 
\usepackage{cleveref}
\crefname{table}{Table}{Tables}
\crefname{equation}{Eq.}{Eqs.}
\crefname{appendix}{App.}{Apps.}
\crefname{section}{Sec.}{Secs.}
\crefname{figure}{Fig.}{Figs.}

\newcommand{\dd}{\text{d}}
\newcommand{\bra}{\langle}
\newcommand{\ket}{\rangle}

\definecolor{customred}{rgb}{0.9, 0.36, 0.36}
\definecolor{customyellow}{rgb}{.92, 0.79, 0.0}
\definecolor{customorange}{rgb}{0.95, 0.55, 0.00}
\definecolor{customgreen}{rgb}{0.17, 0.67, 0.11}
\definecolor{custompurple}{rgb}{0.41, 0.16, 0.68}
\definecolor{custombrown}{rgb}{0.45, 0.25, 0.1}

\begin{document}
\preprint{PITT-PACC-2603}

\title{On the Run from the Dark Side of the Muon}
\author{Pouya Asadi$^{1}$}
\author{Austin Batz$^{2,3}$}
\author{Samuel Homiller$^{4}$}
\author{Tien-Tien Yu$^{2}$}

\affiliation{$^{1}$Department of Physics and Santa Cruz Institute for Particle Physics,\\
University of California Santa Cruz, Santa Cruz, CA 95064}
\affiliation{$^{2}$Institute for Fundamental Science and Department of Physics,\\
University of Oregon, Eugene, OR 97403}
\affiliation{$^{3}$Theoretical Physics Group, Lawrence Berkeley National Laboratory, Berkeley, CA 94720}
\affiliation{$^{4}$Pitt PACC, Department of Physics and Astronomy,\\University of Pittsburgh, Pittsburgh, PA 15260}

\graphicspath{{figs/}}
\begin{abstract}
We present an analysis strategy for probing physics beyond the Standard Model via modifications to the parton distribution functions (PDFs) in a muon beam, which measurably alter the kinematics of all hard processes at a future muon collider.
High-energy muon colliders represent an opportunity to probe new physics using precision measurements and novel search strategies. 
At sufficiently high energies, light particles act as ``constituents'' of the muon described by PDFs. 
As a concrete case study, we apply this framework to an $L_\mu-L_\tau$ gauge boson and demonstrate that, for masses in the range of approximately 50--100 GeV, this indirect PDF-based approach outperforms traditional searches relying on direct gauge boson production. 
These results highlight muon PDF probes as a powerful and promising avenue for beyond the Standard Model physics searches at a future muon collider.
\end{abstract}

\maketitle

\section{Introduction}

The Standard Model (SM) of particle physics provides a remarkably successful description of known fundamental interactions, yet it leaves several key questions unanswered. Among these are the apparent unnaturalness of the electroweak scale, the particle nature of dark matter (DM), and the origin of the observed flavor hierarchies. These open problems suggest the existence of new physics beyond the SM (BSM) and have motivated extensive theoretical and experimental efforts to uncover it.

High-energy colliders play a central role in this endeavor. 
By directly probing short-distance interactions, they offer a uniquely powerful and broadly applicable means of searching for new degrees of freedom. 
In addition to enabling discovery across a wide class of BSM scenarios, collider experiments allow for precise measurements that can discriminate among competing theoretical explanations. 
For example, questions tied to the origin of the electroweak scale and the structure of fundamental interactions are most directly addressed through access to higher energies and luminosities.

Within this context, a muon collider (MuC) has emerged as a compelling candidate for a next-generation high-energy facility~\cite{P5:2023wyd,deBlas:2025gyz}, leading to feasibility studies by the International Muon Collider Collaboration~\cite{InternationalMuonCollider:2025sys} and the US Muon Collider community~\cite{AlAli:2021let,Black:2022cth}. 
As a lepton collider, it 
has a cleaner detector environment than a hadron collider.
Moreover, the muon mass suppresses synchrotron radiation as compared to an electron collider, allowing for circular machines that reach multi-TeV center-of-mass energies ($\sqrt{s}$). 
Unlike protons, muons are not composite particles, which enables a MuC to probe energy scales comparable to those of hadron colliders operating at much higher nominal energies, while potentially offering a more cost-effective path to high luminosity at the energy frontier~\cite{Boscolo:2018ytm}.

The physics potential of a high-energy MuC has been widely explored in recent years.
They have been shown to provide powerful sensitivity to solutions to the electroweak hierarchy problem, models of dark matter, and various other models of BSM physics through direct searches for new particles. See Refs.~\cite{AlAli:2021let, Aime:2022flm, MuonCollider:2022xlm, MuonCollider:2022nsa, Black:2022cth, Accettura:2023ked, InternationalMuonCollider:2025sys, Begel:2025ldu} and references therein for an overview of these studies. 

In this work, we propose a new search strategy at a MuC based on exploiting muon parton distribution functions (PDFs).
While the muon is a fundamental particle, at energies far above its mass, it radiates collinear gauge bosons and fermions that act as constituents with well-defined PDFs.
BSM particles can modify the Dokshitzer–Gribov–Lipatov–Altarelli–Parisi (DGLAP) scale evolution of these PDFs, leading to detectable deviations in kinematic distributions.  Such a strategy was previously explored in the context of dark photons searches at hadron colliders \cite{McCullough:2022hzr}.
Unlike traditional direct search strategies, which rely on measuring hard processes where the BSM particle is produced either on or off-shell, our strategy is to measure the indirect imprint that new physics leaves on the kinematics of SM processes, potentially allowing us to probe BSM coupling values smaller than those accessible to direct searches. The proposed search strategy leverages the unprecedented sensitivity to electroweak physics granted through a high-energy lepton collider.

As an illustrative example, we apply the proposed strategy to probe light muonic forces, focusing on an $L_\mu - L_\tau$ gauge boson with masses around 50--100$\,$GeV.
The phenomenology of such new muonic forces at a high energy MuC \cite{Huang:2021nkl,Dasgupta:2023zrh}, associated beam dump experiments~\cite{Cesarotti:2022ttv, Cesarotti:2023sje}, as well as at other colliders~\cite{CMS:2018yxg, ATLAS:2023vxg,BaBar:2016sci, Belle:2021feg, Belle-II:2024wtd} and neutrino beam experiments~\cite{Altmannshofer:2014cfa,Altmannshofer:2014pba,Altmannshofer:2016jzy}, has been studied before, but we show that our proposal probes a complementary region of parameter space that has remained inaccessible to existing search strategies. 

The remainder of the manuscript is organized as follows. We begin with a review of the DGLAP equations and discuss how BSM physics, using an $L_\mu-L_\tau$ gauge boson as a case study, can impact the SM PDFs. We then present a statistical framework to quantify these effects and conclude with a brief comparison to complementary search strategies.

\section{PDFs and New Physics} \label{sec:pdfs}

The PDFs encode the probabilities for partons to initiate hard interactions while carrying the longitudinal momentum fraction $x$ of the beam. They allow us to factorize the total cross section to produce an inclusive final state as
\begin{equation} \label{eq:sigmatotal}
    \sigma_{\mu^+\mu^-\to F} = \sum_{i,j} 
    \int_0^1\dd x_1 \dd x_2 \ f_{i}(x_1)\bar{f}_{j}(x_2) \,\hat{\sigma}_{ij\to F}\,,
\end{equation}
where the sum is over all combinations of initial state partons $i$ and $j$ carrying longitudinal momentum fractions $x_1$ and $x_2$ of the beams, $F$ is the final state of the hard process initiated by $i$ and $j$, $f_i$ ($\bar{f}_i$) is the PDF of parton $i$ in the (anti-)muon, and $\hat{\sigma}_{ij\to F}$ is the partonic cross section for the initial states $i,j$ to produce $F$.

The PDFs depend on the factorization scale $Q$, which is the characteristic energy scale of the hard interaction. The factorization scale evolution of each PDF depends on the amplitude for parton $A$ to split to partons $B$ and $C$, which is encoded in the splitting function $P^C_{BA}$. The contributions from each $A$ to the evolution of $f_B$ can be resummed into the DGLAP equations, schematically written as
\begin{equation} \label{eq:DGLAPshort}
    \frac{\dd f_B}{\dd\ln Q^2} =  \sum_{A,C}\frac{\alpha_{ABC}}{2\pi} P^C_{BA}\otimes f_A\,,
\end{equation}
where $\otimes$ denotes a Mellin convolution, and $\alpha_{ABC}$ is a coupling (see Refs.~\cite{Chen:2016wkt,Han:2020uid,Han:2021kes, AlAli:2021let, Garosi:2023bvq} for more details). 

Unlike hadron PDFs, muon PDFs can be approximated within perturbation theory.\footnote{The perturbative calculation has finite accuracy and should be supplemented by data-driven fits and matching to fixed-order matrix element calculations once these become available.} Massive partons and electroweak symmetry breaking introduce subtleties that necessitate corrections to the splitting functions in \cref{eq:DGLAPshort}, but the formalism for handling these complications was worked out in Ref.~\cite{Chen:2016wkt}, and we have additional comments in the Supplemental Material \cite{supp}. 
New physics will modify the DGLAP equations and change the PDFs. 

As an example, we consider an $L_{\mu}-L_{\tau}$ gauge boson $Z^\prime$ with the Lagrangian
\begin{align}
\mathcal{L} \supset&-\frac{1}{4}F'^{\alpha\beta}F'_{\alpha\beta}+\frac{1}{2}M_{Z'}^2 Z'^\alpha Z'_\alpha\\
&-g' Z'_{\alpha}  
(\ell_\mu^\dagger\bar{\sigma}^\alpha \ell_\mu + \mu^{c} \sigma^\alpha \mu^{c\,\dagger}
 - \ell_\tau^\dagger\bar{\sigma}^\alpha \ell_\tau - \tau^{c} \sigma^\alpha \tau^{c\,\dagger}\nonumber)\, ,
\end{align}
where $g'$ is the U(1)$_{L_\mu-L_\tau}$ gauge coupling, $M_{Z'}$ is the gauge boson mass, $F'_{\mu\nu}$ is the field strength of $Z'$, $\ell_\mu$ and $\ell_\tau$ are, respectively, the second and third generation left-handed leptons (including the neutrinos), and $\mu^{c\,\dagger}$ and $\tau^{c\,\dagger}$ are, respectively, the second and third generation right-handed leptons. 
The $M_{Z'}$ is generated by an Abelian Higgs mechanism, but we take the radial mode of the dark Higgs to be sufficiently heavy that its impact on phenomenology is negligible. We also neglect any kinetic mixing between U(1)$_{L_\mu-L_\tau}$ and SM U(1)$_{Y}$.

At very high-energies, the $Z'$ can act as a ``parton'' in the muon with its own set of PDFs that contribute to DGLAP via the splittings $f\to f+ Z^\prime$ and $Z^\prime\to f+\bar{f}$, where $f\in\{\mu^\pm, \tau^\pm, \nu_\mu,\nu_\tau\}$. 
While some care is required when defining PDFs for massive particles \cite{Frixione:2023gmf,Frixione:2025wsv,Frixione:2025guf}, our analysis focuses on the region of phase space where $M_{Z^\prime}^2 \ll Q^2$ and emissions are highly collinear, placing us within the regime of validity of the PDF formalism.
The DGLAP evolution of the $Z^\prime$ PDFs back-reacts on each SM PDF, modifying their scale dependence. We compute this effect by modifying the LePDF code package \cite{Garosi:2023bvq} to numerically solve the DGLAP equations with the addition of the $Z^\prime$. 
The log-resummation enabled by numerically solving the DGLAP equations is key to capturing the back-reaction, as leading-log PDF approximations do not include the $Z^\prime$ PDF feeding back into the others. 

\begin{figure}
    \centering
    \includegraphics[width=1.1\linewidth]{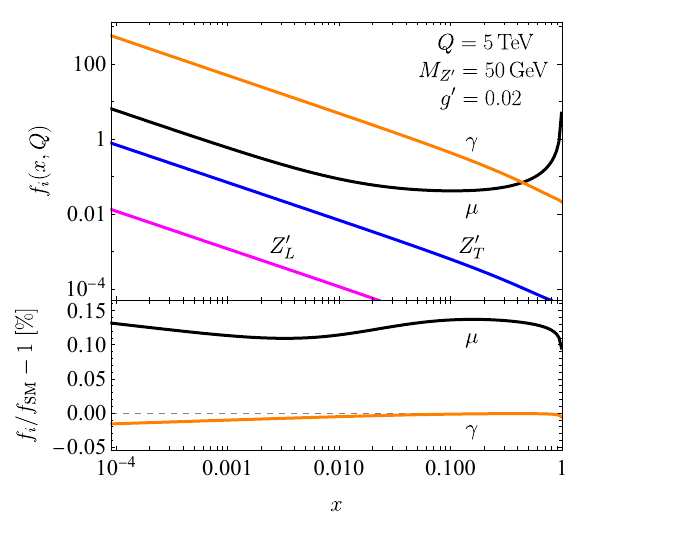}
    \captionsetup{justification=raggedright}
    \caption{{\bf{Upper:}} PDFs for the muon, photon, and $Z^\prime$ (summed over chiral/transverse polarizations) with a particular choice of $Z^\prime$ parameters and factorization scale $Q$. {\bf{Lower:}} percent deviation of the muon and photon PDFs from the SM. The subscript $T$ ($L$) denotes the transverse (longitudinal) polarization. Our calculations also include PDFs for the rest of the SM.}
    \label{fig:pdf}
\end{figure}

The leading effects of the $Z^\prime$ are an \textit{enhancement} to the muon PDF due to a new parton splitting to $\mu^+ + \mu^-$ and a \textit{suppression} of the photon PDF (see \cref{fig:pdf}). 
In a sense, the $Z^\prime$ ``steals probability'' from the photon due to the $\mu\to\mu + Z^\prime$ splitting.  These PDF modifications then modify the final state kinematics of \textit{all} possible hard interactions. In the rest of this work, we present a scheme to probe such deviations and set bounds on the coupling.

\section{Probing PDF Deviations} \label{sec:LmuLtau}

Let $\tau \equiv \hat{s}/s=x_1x_2$ with $\sqrt{\hat{s}}$ the partonic center-of-mass energy. The differential cross section with respect to $\tau$ is 
\begin{equation} \label{eq:dsigmadtau}
    \partial_\tau \sigma \equiv \sum_{i,j} \int_\tau^1 \frac{\dd x}{x} f_i(x)\bar{f}_j\left(\frac{\tau}{x}\right)
    \int \dd \Omega_{\text{fid}} \frac{\dd \hat{\sigma}_{ij}}{\dd \Omega_{\text{lab}}}\,,
\end{equation}
where $\dd \hat{\sigma}_{ij}/\dd \Omega_{\text{lab}}$ is the partonic differential cross section with respect to the lab-frame solid angle, and the integral over $\dd \Omega_{\text{fid}}$ represents integration over angular phase space in a fiducial region of the detector (see the Supplemental Material \cite{supp} for more details). This $\partial_\tau \sigma$ quantity is effectively the experimentally-measured distribution of the final state's invariant mass, and it is the key observable in our analysis. We compute $\partial_\tau\sigma$ by analytically integrating the tree-level amplitudes over phase space, then numerically performing the convolution with the PDFs using $Q=\sqrt{\hat{s}}/2$ as the hard scale.\footnote{The precise definition of $Q$ in terms of $\sqrt{\hat{s}}$ is ambiguous. Varying this definition by a factor of two can create $\mathcal{O}$(10\%) modifications to $\partial_\tau\sigma$ that exceed the deviations we use to probe the $Z^\prime$. We are relying on future higher-precision calculations and data-driven matching being performed to reduce this uncertainty.}


We derive constraints on $g^\prime$ by choosing an ensemble of final states and looking for deviations in their $\tau$ distributions defined in \cref{eq:dsigmadtau}. There are three considerations that influence the choice of final states to analyze:
\begin{enumerate}[label=(\arabic*)]
    \item There is no finalized detector design, so we want final states that can be measured with
    very high efficiency and precision.
    
    \item Matrix element generators such as \textsc{MadGraph5\_aMC@NLO} \cite{Alwall:2014hca,Frederix:2018nkq} do not currently support polarized PDFs, which are necessary due to SM parity violation. Thus, we want two-to-two hard processes with amplitudes that we can analytically compute in perturbation theory.
    
    \item The dominant initial states that produce our chosen final states should be maximally sensitive to PDF deviations. Thus, we want processes dominantly initiated by $\mu^+\mu^-$, $\gamma\gamma$, or $\mu^\pm\gamma$, though we include the small contributions from other tree-level initial states. 
    
\end{enumerate}
The final states that best satisfy these requirements are $\mu^+\mu^-$, $\gamma\gamma$, $\mu^\pm\gamma$, and $e^+e^-$, as we expect the detector to efficiently tag and precisely reconstruct high-energy muons, electrons, and photons.
By comparison, jets, electroweak bosons, and $\tau^\pm$ final states have lower reconstruction efficiency, and neutrinos are not directly measured.  

\begin{figure}
    \centering
    \includegraphics[width=1.0\linewidth]{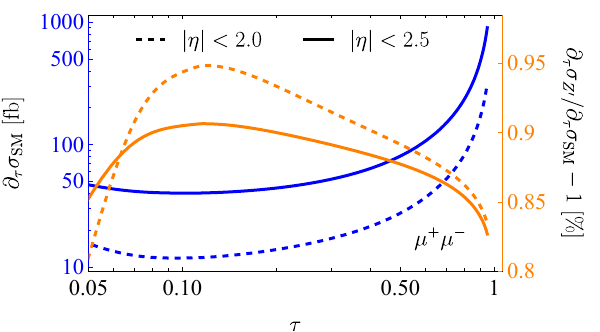}
    \captionsetup{justification=raggedright}
    \caption{{\bf Blue:} SM differential cross section $\partial_\tau \sigma$ for the $\mu^+\mu^-$ final state. {\bf Orange:} percent deviation of $\partial_\tau \sigma$ from the SM for $M_{Z^\prime}=50\,\text{GeV}$ and $g^\prime=0.02$. The solid (dashed) are for $|\eta|<2.5$ $(2.0)$.  \label{fig:dsdtmumu}}
\end{figure}

The effect of the new $Z'$ on the $\partial_\tau \sigma$ distribution is shown in \cref{fig:dsdtmumu} for a benchmark choice of $Z'$ mass and coupling, focusing on a single final-state channel. For this benchmark point, deviations from the SM prediction can reach 
$\mathcal{O}(1\%)$, setting the target level of systematic uncertainty required for this search strategy to be viable. 

We restrict our phase space integration in \cref{eq:dsigmadtau} to the central region of the detector to minimize reconstruction inefficiencies as well as corrections to the PDF approximation that emerge due to forward radiation. 
Thus, both final state particles must be within some pseudorapidity $|\eta|$ range, and we use 2.0 and 2.5 as benchmarks. We also restrict the final state invariant mass to $\geq2\,\text{TeV}$ (\textit{i.e.}~$\tau\geq0.04$) so that power corrections to the PDF approximation due to finite parton masses are suppressed ($M_{Z^\prime}^2/\hat{s}\lesssim\mathcal{O}(10^{-3})$). Finally, we keep $\tau\leq0.95$ to avoid numerical instabilities as the PDF convolution approaches the delta functions of the muon PDFs. Within these ranges, the total cross sections are $\mathcal{O}(10$ -- $100)\,\text{fb}$ for $\mu^+\mu^-$ and $\mathcal{O}(1)\,\text{fb}$ for the others. The $\mu^+\mu^-$ cross sections are particularly large due to strong forward $t$-channel enhancement. 

\section{Statistical Analysis}

Below, we briefly describe our log-likelihood approach to probing deviations in $\partial_\tau \sigma$, reserving additional details for the Supplemental Material \cite{supp}. For a fixed $M_{Z^\prime}$, we constrain $g^\prime$ using the extended log-likelihood for each final state $F$
\begin{equation} \label{eq:lnL}
    \ln L_F = -\bra n_F \ket + n_F \ln \bra n_F \ket 
    + \sum_{i=1}^{n_F} \ln \left(\frac{\partial_\tau \sigma_F}{\sigma_F}\right)\,,
\end{equation}
where the number of events $n_F$ is a Poisson random variable with mean $\bra n_F \ket$, $\partial_\tau\sigma_F/\sigma_F$ is the normalized probability density function for the random variable $\tau$, and the sum over $i$ is a sum over measurements of $\tau$ in each event. The total log-likelihood is the expression in \cref{eq:lnL} summed over the final states $\mu^+\mu^-$, $\gamma\gamma$, $\mu^\pm \gamma$, and $e^+e^-$. This allows us to perform a statistical test to determine the expected maximum value of $g^\prime$ that would result in an experiment consistent with the SM at the 95\% confidence level, which serves as our projected constraint. The inclusion of all final states in the likelihood gives a stronger constraint than any individual process.

To demonstrate the robustness of this shape-driven fitting approach, we include the total integrated luminosity as a source of systematic uncertainty.
To this end, we add the deviations of the luminosity from an expected value of 10$\,\text{ab}^{-1}$ as a nuisance parameter and marginalize over its value. 
We find that the uncertainty on the luminosity has a negligible impact on our constraints, which indicates that the shapes of $\partial_\tau\sigma$ drive the results rather than the normalizations.

Some other sources of experimental uncertainties include misidentifying electrons and photons, as well as beam-induced backgrounds~\cite{MuonCollider:2022ded}. However, these sources can be mitigated through selecting final states that are high-energy and approximately back-to-back. Thus, their contributions are expected to be small. 

Our study can be further refined through future theory developments and an improved understanding of detector design.
Theory uncertainties such as the precise definition of $Q$ and higher-order corrections can be mitigated by a combination of higher-order calculation and data-driven matching of PDFs at low energies.
Our work relies on the accurate identification and precise reconstruction of the final states. 
On the detector side, this will require, for example, large magnetic fields to precisely reconstruct the momenta of multi-TeV muons. 
A realistic detector simulation should account for finite resolution effects.
On the theory side, we will need event generators that include electroweak parton showers to model initial- and final-state radiation \cite{Han:2022laq}, as well as implementing polarized PDFs that faithfully capture the delta functions in the muon PDFs.

\section{Results and Conclusions}

\begin{figure}
    \centering
    \includegraphics[width=1.0\linewidth]{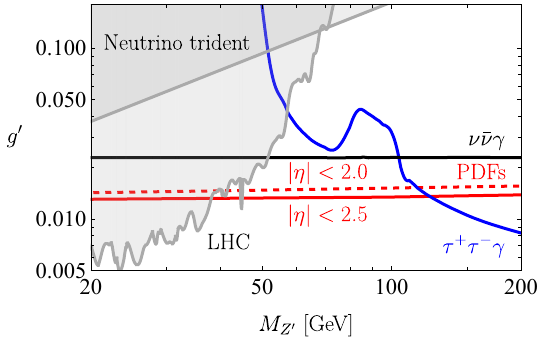}
    \captionsetup{justification=raggedright}
    \caption{Sensitivity to an $L_\mu-L_\tau$ gauge boson. Projected reach obtained using the PDF-based analysis for $|\eta|<2.0$ ({\bf red dashed}) and $|\eta|<2.5$ ({\bf red solid}). Existing constraints ({\bf gray shaded}) include ATLAS~\cite{ATLAS:2023vxg} and CMS~\cite{CMS:2018yxg} searches for $pp\to\mu^+\mu^-\mu^+\mu^-$, where $Z^\prime\to\mu^+\mu^-$ is radiated from a final-state muon in Drell–Yan production, as well as bounds from neutrino trident production~\cite{Altmannshofer:2014cfa,Altmannshofer:2014pba}. Also shown are projected sensitivities from direct searches at a 10~TeV muon collider in the $\nu\bar\nu\gamma$ ({\bf black}) and $\tau^+\tau^-\gamma$ ({\bf blue}) channels.} 
    \label{fig:exclusion}
\end{figure}

In Fig.~\ref{fig:exclusion}, we show the results of our analysis and compare them to other probes for a $L_\mu-L_\tau$ gauge boson. We see that the PDF analysis technique provides the strongest sensitivity for masses between roughly 50 GeV -- 100 GeV. 
The strongest existing constraints above $\sim\! 5~\textrm{GeV}$ come from ATLAS and CMS searches in $p p \to \mu^+\mu^-\mu^+\mu^-$ production, where the $Z' \to \mu^+\mu^-$ is radiated off of a final state muon in Drell--Yan production~\cite{CMS:2018yxg, ATLAS:2023vxg}. 
These set the strongest bounds up to around $60~\textrm{GeV}$.

In addition to direct resonance searches, there are also existing constraints from neutrino trident production, which arise from the scattering of neutrinos in
the Coulomb field of a target nucleus~\cite{Altmannshofer:2014cfa,Altmannshofer:2014pba}. Here, the strongest constraint comes from the measurement of the trident
cross section by the CCFR collaboration~\cite{CCFR:1991lpl}. 

Direct searches for an $L_\mu-L_\tau$ gauge boson are also possible at future lepton colliders. Resonant and associated production at a $\sqrt{s}=3$ TeV muon collider were studied in Refs.~\cite{Huang:2021nkl,Dasgupta:2023zrh}. We recast these results to $\sqrt{s}=10$ TeV in Fig.~\ref{fig:exclusion}, with details of the recasting provided in the Supplemental Material \cite{supp}. 
We see that the $\tau^+\tau^-\gamma$ search has the strongest sensitivity at masses above $\sim 100$ GeV but rapidly loses sensitivity below that, while the sensitivity of the $\nu\bar\nu\gamma$ search is relatively insensitive to $M_{Z'}$. 
Other proposed future colliders, including hadron colliders, can probe a similar part of parameter space to that shown in the figure \cite{Bernardi:2022hny, Francener:2024jra}.


\bigskip


The results of our analysis demonstrate the power of a novel, indirect approach to constraining new physics at a high-energy lepton collider.
In an inversion of traditional electroweak precision studies at lepton colliders, the high-energy of the incoming partons allows for stringent tests of {\em light} (weakly-coupled) new physics. 
Nevertheless, this procedure is systematically improvable: the theoretical uncertainties can be reduced with higher-order, perturbative calculations of the lepton PDFs and the hard scattering process, and final state radiation can be reliably estimated with an electroweak parton shower matched to fixed-order calculations. Our study motivates further developments on all these theoretical aspects.
Our results could further be improved by including additional final states (with appropriate detector simulations), and by including final state angular distributions.
It would also be interesting to study possible synergies of our approach and other searches enabled by forward muon detectors~\cite{Ruhdorfer:2023uea,Li:2024joa,Ruhdorfer:2024dgz, deLima:2025ctj}. 
Our analysis could be straightforwardly applied to various other models of new physics (e.g., muon-philic scalars, axion-like particles, or heavy neutral leptons). 
Our hope, however, is that this study motivates using novel ideas of how beyond the Standard Model physics can be unearthed with the unique capabilities of a high-energy muon collider.

\section{Acknowledgments}

We thank Christian Bauer, Lance Dixon, Tao Han, Simon Knapen, Patrick Meade, Michael Peskin, Dean Robinson, and Natalia Toro for helpful discussions. 
The work of PA is supported in part by the US Department of Energy grant number DE-SC0010107.
T-TY is supported in part by the U.S. Department of Energy under Grant Number DE-SC0011640. 
The work of AB is supported by the US Department of Energy, Office of Science Graduate Research program under contract number DE-SC0014664.


\bibliography{refs}
\bibliographystyle{utphys}

\onecolumngrid
\clearpage
\begin{center}
\textbf{\large Supplemental Materials: On the Run from the Dark Side of the Muon}
\end{center}
\setcounter{equation}{0}
\setcounter{figure}{0}
\setcounter{table}{0}
\setcounter{section}{0}
\setcounter{subsection}{0}
\setcounter{page}{1}
\makeatletter
\renewcommand{\theequation}{S\arabic{equation}}
\renewcommand{\thefigure}{S\arabic{figure}}
\renewcommand{\bibnumfmt}[1]{[S#1]}
\renewcommand{\thesection}{S\arabic{section}}
\renewcommand{\thesubsection}{S\arabic{subsection}}

\onecolumngrid


\section{Computing Parton Distribution Functions}

\subsection{\texorpdfstring{DGLAP Evolution in L\MakeLowercase{e}PDF}{DGLAP Evolution in LePDF}}

Here, we briefly review DGLAP evolution and its implementation in LePDF \cite{Garosi:2023bvq}, which we use to numerically compute each parton's PDF in perturbation theory. Much of the formalism for applying DGLAP to electroweak processes was worked out in Ref.~\cite{Chen:2016wkt}. The DGLAP equations are a system of coupled differential equations:
\begin{equation} \label{eq:DGLAP}
    Q^2\frac{\dd f_B}{\dd Q^2}
    =
    P_B^v f_B
    + \sum_{A,C} \frac{\alpha_{ABC}}{2\pi}
    \widetilde{P}_{BA}^C \otimes f_A 
    + \frac{v^2}{16\pi^2Q^2} 
    \sum_{A,C} \widetilde{U}^C_{BA}\otimes f_A\,.
\end{equation}
The PDF $f_B$ is a function of the longitudinal momentum fraction $x$ of parton $B$ in the beam, as well as the factorization scale $Q$ (the characteristic momentum scale of the hard interaction, which we treat as $\sqrt{\hat{s}}/2$ in this work). The Altarelli-Parisi splitting function $P_{BA}^C$ \cite{Altarelli:1977zs} can be written in terms of the amplitude of the one-to-two splitting process $A\to B+C$:
\begin{equation}
    |\mathcal{M}_{A\to B+C}|^2 = 8\pi \alpha_{ABC}\frac{p_T^2}{x(1-x)} P^C_{BA}(x)\,,
\end{equation}
where $\alpha_{ABC}$ is the appropriate coupling, and $p_T$ is the transverse momentum of $B$. The tilde on $\widetilde{P}_{BA}^C$ denotes the mass correction
\begin{equation} \label{eq:tilde}
    \widetilde{P}_{BA}^C = \left(  
    \frac{p_T^2}{\tilde{p}_T^2}
    \right)^2
    P^C_{BA}\,,
\end{equation}
where
\begin{equation}
\tilde{p}_T^2 \equiv p_T^2 + x \,m_C^2 + (1-x)\,m_B^2 - x\,(1-x)\,m_A^2\,.
\end{equation}
One makes the replacement $p_T^2\to Q^2$ when inserting the splitting functions into \cref{eq:DGLAP}. The splitting functions are convolved with the PDFs via the usual Mellin convolution
\begin{equation}
    (f\otimes g)(x) = \int_x^1\frac{\dd z}{z} f(z)g\!\left( \frac{x}{z} \right).
\end{equation}
The $U_{BA}^C$ functions are known as ultra-collinear splitting functions and are important in the electroweak-broken phase due to the massive $W^\pm$ and $Z$ splitting amplitudes containing $1/p_T^4$ behavior in addition to the $1/p_T^2$ behavior seen in the massless case (see Ref.~\cite{Chen:2016wkt} for details). The ultra-collinear contributions depend on the Higgs vacuum expectation value (vev) $v$. The tilde on $\widetilde{U}_{BA}^C$ has an analogous definition to \cref{eq:tilde}. Finally, $P_B^v$ is a virtual correction that cancels infrared divergences in the limit $x\to 1$. For example, the electron splitting function $P_{ee}^\gamma$ in pure quantum electrodynamics is sometimes written \cite{Peskin:1995ev}
\begin{equation} \label{eq:electronsplitting}
    P_{ee}^\gamma(x) = \frac{1+x^2}{(1-x)_+}
    + \frac{3}{2} \delta(1-x)\,,
\end{equation}
with the usual plus function
\begin{equation}
    \int_x^1\dd z \frac{f(z)}{(1-z)_+} = \int_x^1\dd z\frac{f(z) - f(1)}{1-z}\,.
\end{equation}
In the convention used by LePDF, the contribution of the delta function term in \cref{eq:electronsplitting} is contained in the virtual correction.

A key difference between PDFs in the muon and proton is that the boundary conditions to solve the system of differential equations in \cref{eq:DGLAP} are based on perturbative physics. For the proton, one can evolve the PDFs between scales within perturbation theory, but the boundary conditions at the initial scale must be tuned to measurements. In the muon case, the boundary condition is fixed by setting $Q$ to the muon mass:
\begin{equation}
    f_\mu(x,Q=m_\mu) = \delta(1-x)\,.
\end{equation}
Loosely, this corresponds to the intuitive statement that when the muon's energy equals the muon mass, it is composed entirely of the muon. The other needed boundary conditions are that when the scale is below the mass of a particular parton, that parton's PDF vanishes. These conditions uniquely specify the solution to \cref{eq:DGLAP}.

In practice, there are non-perturbative corrections due to quantum chromodynamics. 
These corrections can be quantified by a data-driven fit to determine the PDFs at low energies, which can then be evolved using DGLAP to higher energies, as is done for the proton PDF at the LHC.
Regardless, one can determine the boundary condition for solving the DGLAP equations. In LePDF, this subtlety is handled by setting the gluon PDF to zero for $Q<700\,\text{MeV}$.
We expect these corrections to be suppressed by the ratio of the confinement scale to $Q$.

The delta function component of the muon PDF within the muon introduces some subtleties in the numerical calculations. In LePDF, the final $x$ bin is used to contain the delta function contribution, and the value of the PDF in this bin is determined by enforcing the momentum conservation sum rule
\begin{equation}
    \sum_i \int_0^1 \dd x \, x \, f_i(x) = 1
\end{equation}
at each step of DGLAP evolution. Numerically integrating functions involving the muon PDF requires some care, as attempting to do so by interpolating the PDF between $x$ bins can introduce significant error when interpolating between the last and next-to-last bins. One is therefore restricted to Riemann sums whenever evaluating the argument of the PDF near $x=1$. This introduces some complication when evaluating integrals involving PDFs of both the muon and anti-muon such as those in \cref{eq:dsigmadtau}. When performing Riemann sums for integrals of the form
\[
    \int_\tau^1\frac{\dd x}{x} f_i(x)\bar{f}_j\left(\frac{\tau}{x}\right)\,g\left(x,\frac{\tau}{x}\right)\,
\]
(where $g$ is an arbitrary function), the argument of $\bar{f}_j$ does not align with the pre-determined $x$ bins tabulated by LePDF, so $\bar{f}_j$ must be interpolated between bins. This becomes problematic as $x\to\tau$ and the argument of $\bar{f}_j$ approaches one. This problem can be remedied by splitting the integral into two pieces:
\begin{equation}
    \int_\tau^1\frac{\dd x}{x} f_i(x)\bar{f}_j\left(\frac{\tau}{x}\right)\,g\left(x,\frac{\tau}{x}\right) = \int_{\sqrt{\tau}}^1\frac{\dd x}{x} f_i(x)\bar{f}_j\left(\frac{\tau}{x}\right)\,g\left(x,\frac{\tau}{x}\right) +\int_{\sqrt{\tau}}^1\frac{\dd x}{x} f_i\left(\frac{\tau}{x}\right)\bar{f}_j(x)\,g\left(\frac{\tau}{x},x\right) .
\end{equation}
This way, one can use the pre-determined $x$ bins for $f_i$ in the first term and $\bar{f}_j$ in the second term in the limit where the argument approaches one. As $x\to\sqrt{\tau}$, the argument of the interpolated function (\textit{i.e.}~$\tau/x$) remains strictly less than one (since we impose $\tau \leq 0.95$ in our calculation), which is where the interpolation is valid.

\subsection{\texorpdfstring{Modifications to L\MakeLowercase{e}PDF}{Modifications to LePDF}}

To include the $Z^\prime$ PDFs, we must add both the transverse polarizations $f_{Z_\pm^\prime}$ and longitudinal polarization $f_{Z_L^\prime}$. 
As further commented on below, we also need mixed PDFs to account for interference between the $Z^\prime$ and both the photon and $Z$ for each polarization: $f_{\gamma_\pm / Z^\prime_\pm}$, $f_{Z_\pm / Z^\prime_\pm}$, and $f_{Z_L / Z^\prime_L}$. 
We assume that the radial mode of the scalar that breaks U(1)$^\prime$ has negligible impact, so we do not include PDFs for this state or its interference with $Z_L^\prime$. 
The Higgs couplings to leptons are completely negligible in this work, so we also do not include a $h/Z_L^\prime$ mixed PDF. 
The $Z^\prime$ has ultra-collinear splitting functions completely analogous to the $Z$, but their contributions to DGLAP in \cref{eq:DGLAP} must be weighted by the vev of the scalar that breaks U$(1)^\prime$ rather than the SM Higgs vev. 
Otherwise, the various splitting functions that involve the $Z^\prime$ can be almost read off of the corresponding $Z$ splitting functions enumerated in Ref.~\cite{Garosi:2023bvq} with appropriate replacements of charges and couplings. 
The virtual corrections are similarly adapted.

A subtlety in adapting LePDF to an $L_\mu-L_\tau$ model is that the coupling violates the lepton universality present in the SM. In its unadapted form, LePDF takes advantage of this symmetry by combining the anti-muon neutrino with the anti-electron neutrino and the anti-muon with the positron. Our $Z^\prime$ breaks this symmetry, which requires us to separate out these flavors.

Another subtlety is that LePDF evolves the PDFs in two phases: below and above the electroweak scale. An advantage of this approach is that the SM with the $W^\pm$ and $Z$ omitted preserved parity and charge symmetries, so the PDFs of different chiral/polarization/anti-particle states need not be separated below the electroweak scale. Moreover, the neutrinos in the SM only arise as partons in the muon due to splittings of the $W^\pm$ and $Z$, so the neutrino PDFs can be omitted below the electroweak scale. For our case, when $M_{Z^\prime}$ is below the electroweak scale, the $\nu_\mu$ and $\nu_\tau$ become active at $Q=M_{Z^\prime}$ since the $Z^\prime$ can split to them. 

For these reasons, we must introduce PDFs for the following six partons below the electroweak scale: 
\[Z^\prime_T,Z^\prime_L,\gamma_T/Z^\prime_T,\nu_\mu,\nu_\tau,\mu^+,\]
where the neutrino/anti-neutrino and transverse polarizations are combined. Above the electroweak scale, we introduce PDFs for the following eleven partons:
\[Z^\prime_\pm,Z^\prime_L,\gamma_\pm/Z^\prime_\pm,Z_\pm/Z^\prime_\pm,Z_L/Z^\prime_L,\bar{\nu}_\mu,\mu^+_{L/R},\]
as now each chirality/polarization/anti-particle/lepton flavor must be separated. Each of these PDFs has its own DGLAP evolution equation.

\section{Further Details on the Invariant Mass Distributions}
\subsection{Fiducial Phase Space}
\label{sec:fiducial}

In this section, we discuss in detail the semi-analytic approach to calculating $\partial_\tau\sigma$ in \cref{eq:dsigmadtau}. 
First, we differentiate \cref{eq:sigmatotal} with respect to $\tau=x_1x_2$ and the lab-frame solid angle:
\begin{equation}
    \frac{\dd^2 \sigma}{\dd\tau\,\dd\Omega_{\text{lab}}}
    =
    \sum_{i,j}\int_\tau^1\frac{\dd x}{x}f_i(x)\bar{f}_j\left(\frac{\tau}{x}\right)\frac{\dd \hat{\sigma}_{ij}}{\dd\Omega_{\text{lab}}}\,.
\end{equation}
We consider the lab-frame solid angle because our angular phase space integral must be over some region of the detector, and the hard process is boosted relative to the detector with rapidity $\hat{y}=\ln(x_1/x_2)/2=\ln(x^2/\tau)/2$. 
The most convenient angular coordinate is the rapidity $y=\tanh^{-1}(\cos(\theta))$, where $\theta$ is the polar angle of one of the final states in the lab frame. For a massless two-to-two process, we can write the partonic differential cross section as 
\begin{equation}
   \frac{\dd \hat{\sigma}_{ij}}{\dd\Omega_{\text{lab}}} = \frac{1}{64\pi^2 x_1x_2 s}\frac{\cosh^2y}{\cosh^2(y-\hat{y})} |\mathcal{M}_{ij}|^2\,,
\end{equation}
where $\mathcal{M}_{ij}$ is the partonic matrix element for the process. Integrating this over the fiducial region, we have 
\begin{equation} \label{eq:fidintegral}
    \int\dd\Omega_{\text{fid}}\frac{\dd \hat{\sigma}_{ij}}{\dd\Omega_{\text{lab}}}
    =
    \frac{2\pi}{1+\delta_{F_1F_2}}
    \frac{1}{64\pi^2 x_1x_2 s}
    \int_{\mathcal{D}}\dd y \frac{|\mathcal{M}_{ij}|^2}{\cosh^2(y-\hat{y})},
\end{equation}
where $\delta_{F_1F_2}=1$ if the final states are identical, and $\mathcal{D}$ is the domain of integration
\begin{equation} \label{eq:integraldomain}
    \mathcal{D} = \{y \, |\,  \max(-y_{\text{max}},2\hat{y}-y_{\text{max}}) 
    \leq
    y \leq \min(y_{\text{max}},2\hat{y}+y_{\text{max}})\}\,,
\end{equation}
within which \textit{both} final states have $|\eta|\leq y_{\text{max}}$. 
Our benchmarks of $y_{\text{max}}=2.0$ and $y_{\text{max}}=2.5$ are, respectively, more conservative and more optimistic about the angular region where measurements will have high efficiency and precision.
The domain of integration depends on $x_1$ and $x_2$, so in order to maintain analytical expressions for as much of the calculation as possible, we define the \textit{indefinite} integral
\begin{equation} \label{eq:Ihat}
    \hat{I}_{ij} = \frac{1}{64\pi^2 x_1x_2 s}
    \int \dd y \frac{|\mathcal{M}_{ij}|^2}{\cosh^2(y-\hat{y})}.
\end{equation}
For example, the annihilation of a fermion with an anti-fermion of opposite chirality to $\gamma\gamma$ gives
\begin{equation}
    \hat{I}_{ij} = \frac{4\alpha^2Q_f^4}{x_1x_2s}\left(  
    y-\frac{1}{2}\tanh(y-\hat{y})
    \right),
\end{equation}
where $Q_f$ is the fermion's electric charge, and $\alpha$ is the electromagnetic coupling. Thus, one can evaluate the definite integral in \cref{eq:fidintegral} by applying the Fundamental Theorem of Calculus with the lower and upper $y$ bounds in \cref{eq:integraldomain}, and numerics only come in during the evaluation of the $x$ integral.

\subsection{Interfering Initial States}

When a particular final state can be produced in a hard process by initial states with identical quantum numbers, the total cross section contains interference terms. For example, the $\mu^-\gamma$ final state can be produced by the initial states $\mu^-\gamma$, $\mu^- Z$, and $\mu^- Z^\prime$ (as well as the further-suppressed $\nu_\mu W^-$ and negligible $\mu^-h$, but let us put those aside for now). Schematically, there is interference between the diagrams

\begin{figure}[H]
    \centering
    \begin{minipage}[t]{0.17\linewidth}
    \includegraphics[width=\linewidth]{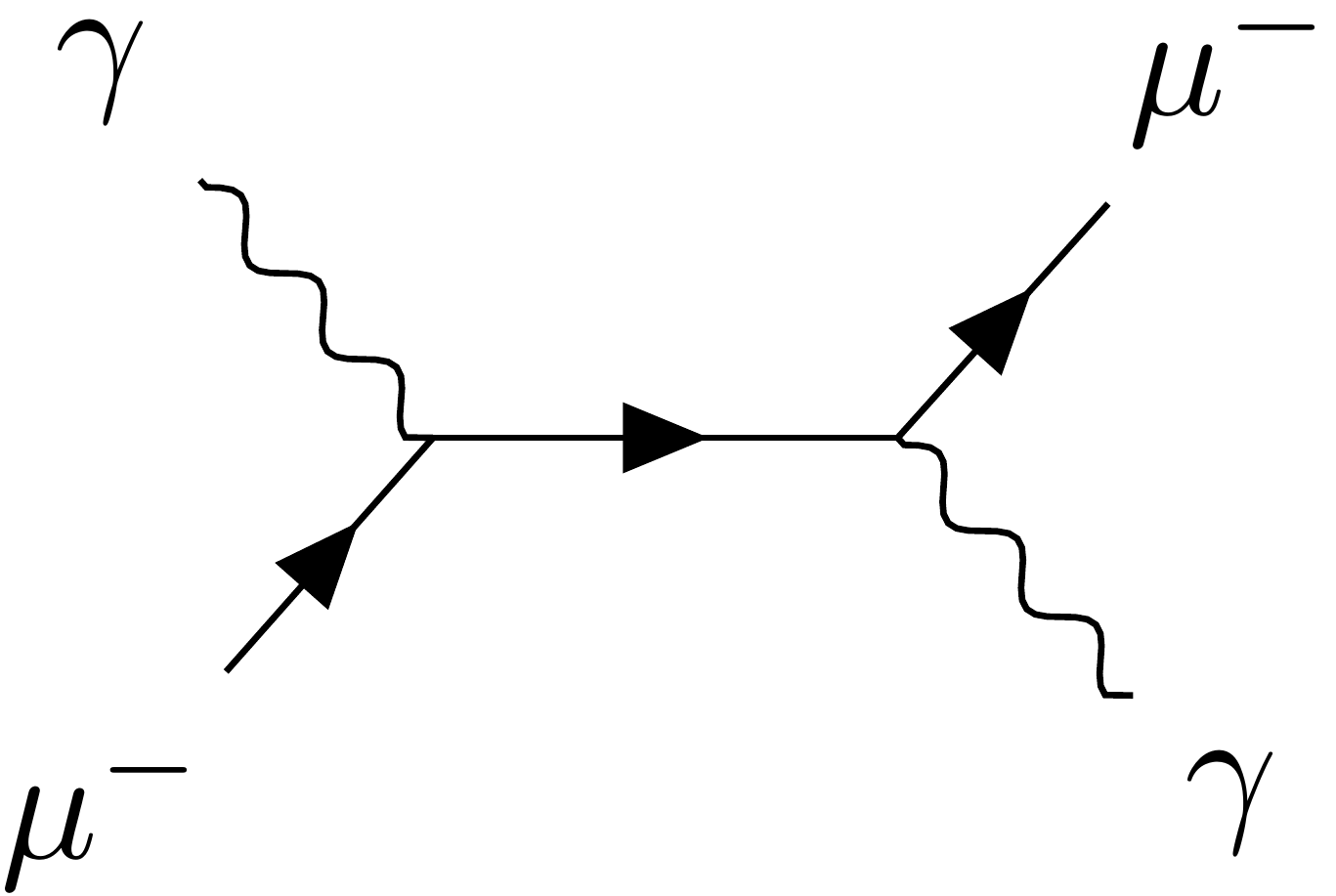}
    \end{minipage}
    \hspace{0.05\textwidth} 
    \begin{minipage}[t]{0.17\linewidth}
    \includegraphics[width=\linewidth]{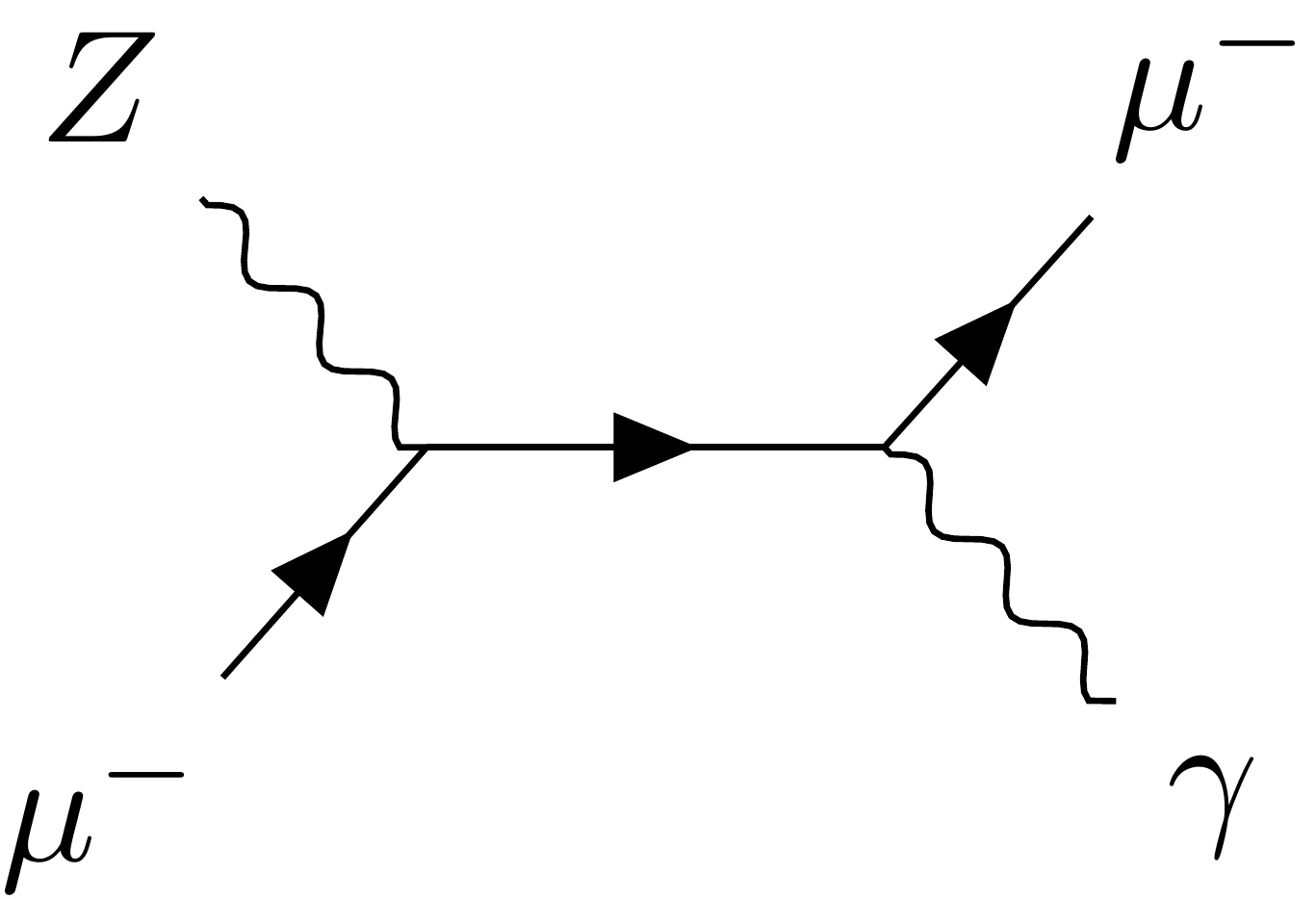}
    \end{minipage}
    \hspace{0.05\textwidth} 
    \begin{minipage}[t]{0.17\linewidth}
    \includegraphics[width=\linewidth]{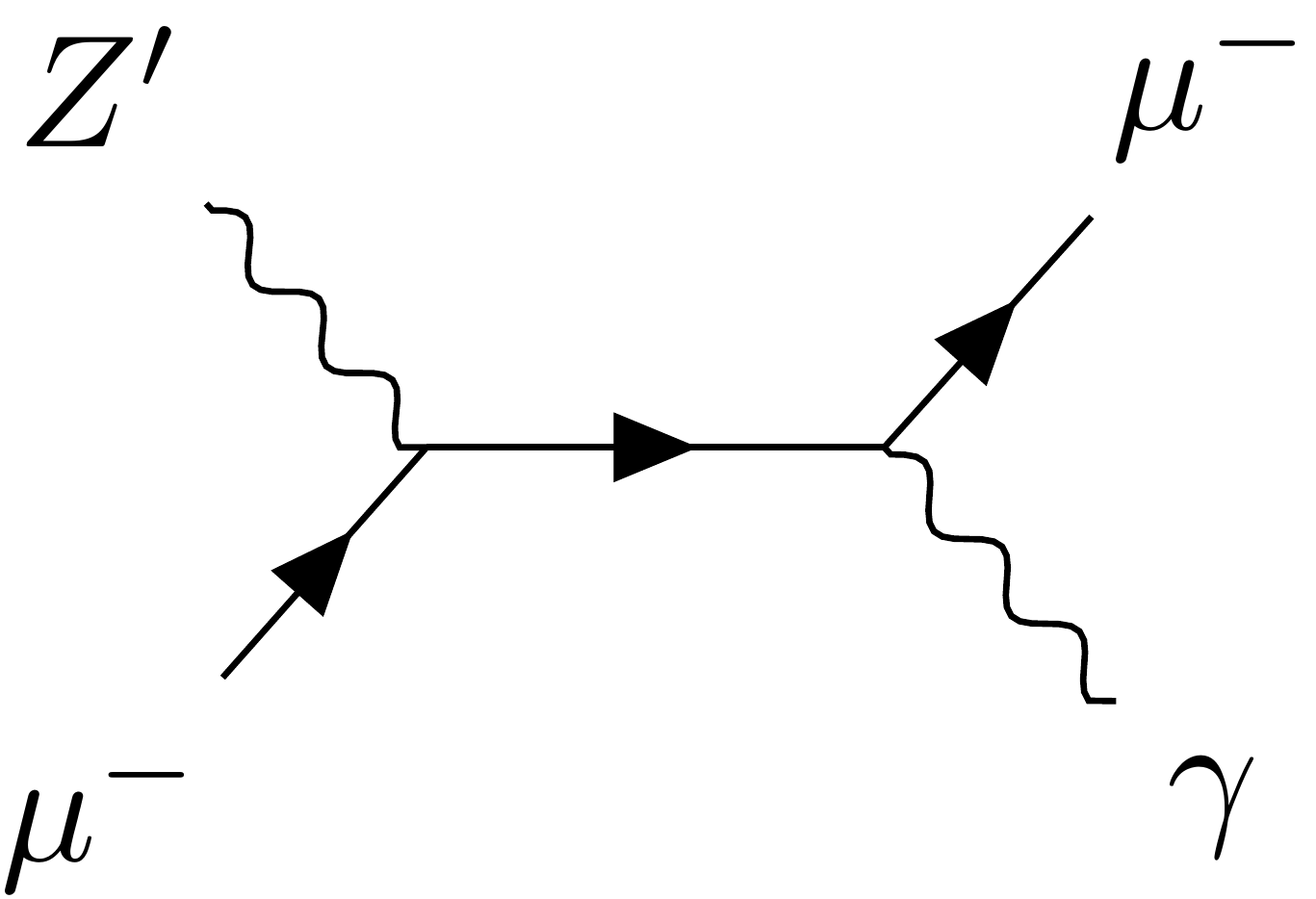}
    \end{minipage}
    \label{fig:placeholder}
\end{figure}

\noindent for each transverse polarization of the initial $\gamma$, $Z$, and $Z^\prime$ and the longitudinal polarizations of the $Z$ and $Z^\prime$. This effect is captured in the PDF approach by DGLAP evolution of mixed states with their own separate PDFs  ($f_{\gamma_\pm/Z_\pm}$, $f_{\gamma_\pm/Z^\prime_\pm}$, $f_{Z_L/Z^\prime_L}$, \textit{etc}.) \cite{Chen:2016wkt,Marzocca:2024ica}. In such cases, the integrand in the $\partial_\tau\sigma$ calculation contains not only terms with factors of the form $f_i\bar{f}_j |\mathcal{M}_{ij}|^2$, but also some like 
\begin{equation}
    2f_{i_1/i_2}\bar{f}_j \text{Re}(\mathcal{M}_{i_1j}^\ast\mathcal{M}_{i_2j})\,,
\end{equation}
where $i_1$ and $i_2$ are initial states that interfere with each other, the two matrix elements correspond to each diagrams with each initial state, and one takes two times the real part of the product due to adding the product of matrix elements to its complex conjugate.

For the $\mu^+\mu^-$ final state, there are many interference terms due to diagrams of the form

\begin{figure}[H]
    \centering
    \includegraphics[width=0.2\linewidth]{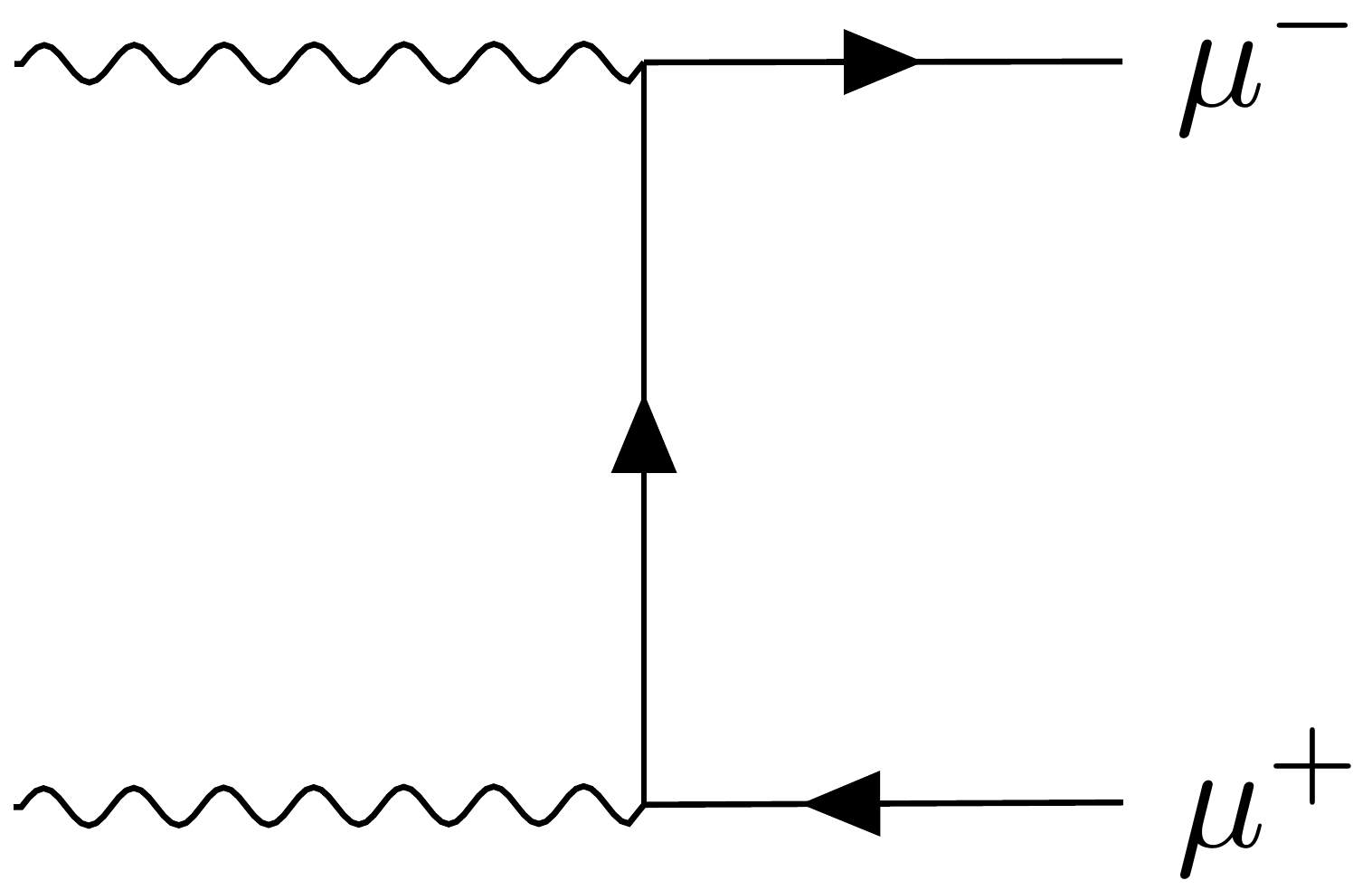}
\end{figure}

\noindent with combinations of photon, $Z$, and $Z^\prime$ initial states. In this case, one must also include interference terms in which both PDFs correspond to mixed states. For the purposes of this work, these effects are very small in practice, but we include all initial states that can produce each final state at tree level for completeness.

\subsection{Additional Invariant Mass Distributions}

\begin{figure}[t]
    \centering
    \begin{minipage}[t]{1.0\linewidth}
    \includegraphics[width=0.49\linewidth]{figs/dsdtmumu}
    \includegraphics[width=0.49\linewidth]{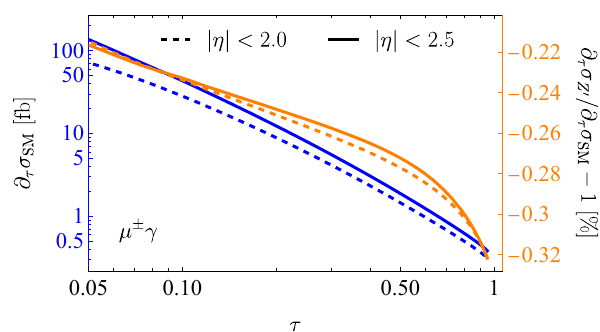}
    \includegraphics[width=0.49\linewidth]{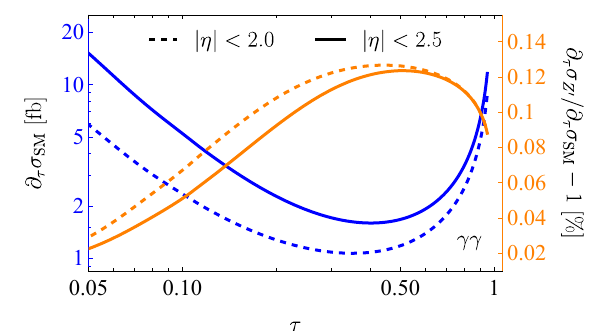}
    \includegraphics[width=0.49\linewidth]{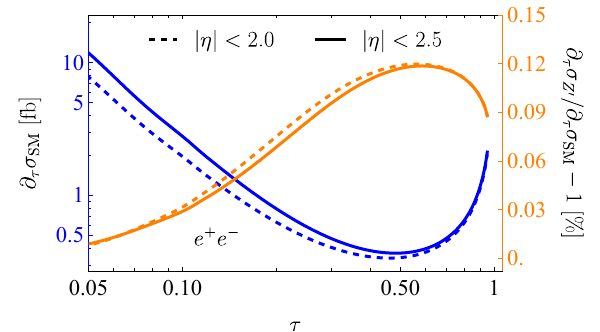}
    \end{minipage}
    \captionsetup{justification=raggedright,width=0.9\linewidth}
    \caption{{\bf{Blue:}} SM differential cross section $\partial_\tau \sigma$ for each final state we consider. {\bf{Orange:}} percent deviation of $\partial_\tau \sigma$ from the SM for $M_{Z^\prime}=50\,\text{GeV}$ and $g^\prime=0.02$.  
    \label{fig:dsdtall}}
\end{figure}

In \cref{fig:dsdtall}, we show $\partial_\tau \sigma$ distributions for each of our final states, as well as representative deviations of the $Z^\prime$ model from the SM (the $\mu^+\mu^-$ plot is repeated from the main text). The presence of the $Z^\prime$ increases the total cross sections except in the $\mu^\pm$ case. This is because the $Z^\prime$ \textit{enhances} the muon PDF for $x<1$ via back-reaction in DGLAP (because it can split to $\mu^+\mu^-$ pairs), while the $Z^\prime$ ``steals probability'' away from the photon. The $\mu^\pm\gamma$ final state is the most sensitive to the photon PDF because, unlike the others, it cannot be produced by the dominant $\mu^+\mu^-$ initial state. If these sorts of kinematic deviations from the SM are discovered, the fact that different final states reveal deviations in different parton PDFs may be useful in determining which partons the new physics couples to, which would help distinguish a $Z^\prime$ coupling to $L_\mu-L_\tau$ from some other theory.

It is no coincidence that the shapes of the $\partial_\tau\sigma$ distributions for $\mu^+\mu^-$, $\gamma\gamma$, and $e^+e^-$ bear a visual resemblance to the muon PDF itself (at least near $\tau=1$), nor that the $\mu^\pm\gamma$ distribution resembles the photon PDF. In the cases where the dominant initial state is $\mu^+\mu^-$, the largest contribution to the PDF convolution in \cref{eq:dsigmadtau} comes from the delta function component of either the $\mu^-$ or $\mu^+$ PDF convolved with the PDF of the other. Likewise, the dominant contribution for $\mu^\pm\gamma$ comes from the delta function component of the $\mu^-$ or $\mu^+$ PDF convolved with the photon PDF. 

For each final state, the absolute value of the fractional deviation from the SM tends to be larger for $|\eta|<2.0$ than for $|\eta|<2.5$, which indicates that the SM rate increases more quickly as a function of $|\eta|$ than the deviations due to the $Z^\prime$. 
Since both $\partial_\tau \sigma_{\text{SM}}$ and $\partial_\tau \sigma_{Z^\prime}$ are forward-enhanced, the denominator of the ratio
\begin{equation}
\frac{\partial_\tau \sigma_{Z^\prime} - \partial_\tau \sigma_{\text{SM}}}{\partial_\tau \sigma_{\text{SM}}}
\end{equation}
receives a forward enhancement that may be partially canceled between the two terms in the numerator.

\section{Further Details on the Statistical Analysis}

Here, we provide a more comprehensive description of the statistical procedure we use to derive constraints on $g^\prime$. A more detailed and general expression of the extended log-likelihood in \cref{eq:lnL} is
\begin{equation} \label{eq:lnLDetailed}
    \ln L(g^\prime,\nu) = -\frac{(\nu-\mu_\nu)^2}{2\sigma_\nu^2}
    + \sum_F \left(-\bra n_F(g^\prime,\nu) \ket + n_F \ln \bra n_F(g^\prime,\nu) \ket 
    + \sum_{i=1}^{n_F} \ln \left(\frac{\partial_\tau\sigma_F(g^\prime,\nu)}{\sigma_F(g^\prime,\nu)}\right)
    \right)\,,
\end{equation}
where $\nu$ is a nuisance parameter we use to handle systematic uncertainty in the integrated luminosity, $\mu_\nu$ is the expectation value $\nu$, $\sigma_\nu$ is the uncertainty in $\nu$, the $F$ subscript labels the final states (which are summed over), and we have explicitly written the expectation value of the number of events and the $\sigma_F$ as functions of $g^\prime$ and $\nu$. In our case, $\sigma_F$ is independent of $\nu$, as $\nu$ only impacts the total integrated luminosity. 
Thus, we can rewrite the expectation value of the number of events in the simple form
\begin{equation}
    \bra n_F(g^\prime,\nu) \ket = \mathcal{L}_0(1+\nu)\,\sigma_F(g^\prime)\,,
\end{equation}
where $\mathcal{L}_0=10\,\text{ab}^{-1}$ is the expected luminosity, and $\mu_\nu=0$ by this definition of $\nu$. The simplicity of our nuisance parameter allows us to analytically profile over its value (\textit{i.e.}~find the value that maximizes $\ln L$ for an arbitrary $g^\prime$):
\begin{equation}
    \hat{\hat{\nu}} = \frac{1}{2}\left(
    -1 - \sigma_\nu^2 \, \mathcal{L}_0 \sum_F \sigma_F(g^\prime)
    +
    \sqrt{4\,\sigma_\nu^2\sum_F n_F + \left( 1 - \sigma_\nu^2 \, \mathcal{L}_0 \sum_F \sigma_F(g^\prime) \right)^2}
    \right),
\end{equation}
which we use to marginalize over $\nu$. We find that the chosen value of $\sigma_\nu$ has negligible impact on our results. 

The log-likelihood allows us to conduct a statistical test to determine the range of $g^\prime$ that would cause the experiment to be consistent with the SM at a particular confidence level. The appropriate test statistic is \cite{Cowan:2010js}
\begin{equation}
    q = 
\begin{cases}
    -2\ln \frac{L(g^\prime,\hat{\hat{\nu}}(g^\prime))}{L(0,\hat{\hat{\nu}}(0))}, & \hat{g}^\prime < 0, \\
    -2\ln \frac{L(g^\prime,\hat{\hat{\nu}}(g^\prime))}{L(\hat{g}^\prime,\hat{\nu})}, & 0\leq \hat{g}^\prime < g^\prime, \\
    \mathmakebox[\widthof{$-2\ln \frac{L(g^\prime,\hat{\hat{\nu}}(g^\prime))}{L(\hat{g}^\prime,\hat{\nu})}$}][r]{0}, & g^\prime < \hat{g}^\prime,
\end{cases}
\end{equation}
where a single hat denotes the value of the argument that unconditionally maximizes $\ln L$, and the double hat denotes the profiled value. This test statistic accounts for the fact that a negative $\hat{g}^\prime$ is unphysical and that we do not consider values of $g^\prime$ below $\hat{g}^\prime$ (\textit{i.e.}~this is a one-tailed test). We estimate the profiled value of $\ln L$ as a function of $g^\prime$ by performing an ensemble of toy experiments, sampling each $n_F$ from a Poisson distribution assuming SM cross sections, then sampling $\tau$ for each event and final state assuming SM $\partial_\tau\sigma_F$ distributions. We then average the value of $\ln L$ over the ensemble of experiments for several values of $g^\prime$. This yields a parabolic function of $g^\prime$, the maximum of which is located at $\hat{g}^\prime$. We can then define the ``standard deviation'' $\sigma_{g^\prime}$ via
\begin{equation}
    q = 
\begin{cases}
    \left( \frac{g^\prime}{\sigma_{g^\prime}} \right)^2 - \frac{2g^\prime\hat{g}^\prime}{\sigma_{g^\prime}^2}, & \hat{g}^\prime < 0, \\
     \mathmakebox[\widthof{$\left( \frac{g^\prime}{\sigma_{g^\prime}} \right)^2 - \frac{2g^\prime\hat{g}^\prime}{\sigma_{g^\prime}^2}$}][r]{\left( \frac{g^\prime-\hat{g^\prime}}{\sigma_{g^\prime}} \right)^2},& 0\leq\hat{g}^\prime <g^\prime, \\
     \mathmakebox[\widthof{$\left( \frac{g^\prime}{\sigma_{g^\prime}} \right)^2 - \frac{2g^\prime\hat{g}^\prime}{\sigma_{g^\prime}^2}$}][r]{0} , & g^\prime < \hat{g}^\prime. 
\end{cases}
\end{equation}
Our projected constraint at the 95\% confidence level is then
\begin{equation}
    g^\prime_{95} = \max(0,\hat{g}^\prime) + \sigma_{g^\prime}\,Z_{95}\,,
\end{equation}
where $Z_{95}\simeq1.645$ is the significance corresponding to a 95\% confidence interval for a one-tailed test, and the maximum is to enforce physicality of the estimate. 

\begin{figure}
    \centering
    \begin{minipage}[t]{\linewidth}
    \includegraphics[width=0.49\linewidth]{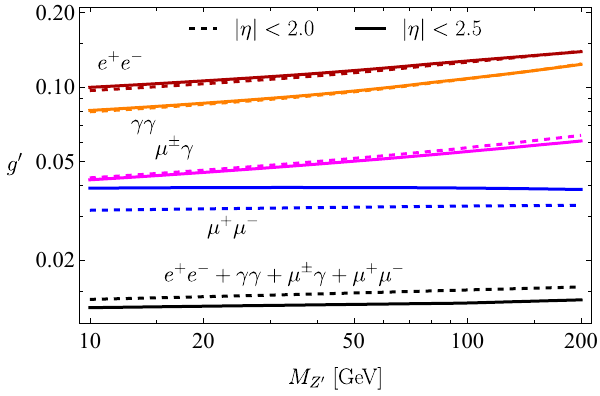}
    \includegraphics[width=0.49\linewidth]{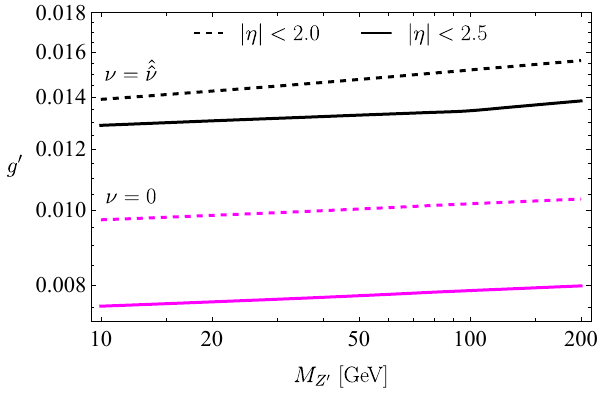}
    \captionsetup{width=\linewidth,justification=raggedright}
    \caption{
    Projected constraints on $g^\prime$ using different statistical procedures. {\bf Left:} limits using each final state individually and the combination of all final states. The combination of all final states provides a much stronger constraint than any one individually. {\bf Right:} limits using all final states with marginalization over total luminosity (black, $\nu=\hat{\hat{\nu}}$) and without marginalization (magenta, $\nu=0$). Accounting for this systematic uncertainty strongly influences the results. 
    \label{fig:finalStateLimits}
    }
    \end{minipage}
\end{figure}

The combination of different final states plays a key role in the strength of the constraint. As shown in the left panel of \cref{fig:finalStateLimits}, using all four final states is much more powerful than using any one individually. This observation further highlights the fact that our approach can be systematically improved by including the wide array of potential final states that a real experimental analysis would be able to exploit given knowledge of the detector. The systematic uncertainty in the total luminosity also strongly impacts the results, as shown in the right panel of \cref{fig:finalStateLimits}. This shows that our analysis is largely insensitive to deviations from the SM in the total cross sections, which is corroborated by the fact that our results are approximately independent of $\sigma_\nu$.

The shapes of the fractional $\partial_\tau\sigma$ deviations from the SM (shown as the orange curves in \cref{fig:dsdtall}) are what our statistical analysis is most sensitive to. Our marginalization over the total luminosity causes us to lose sensitivity to deviations in the total rates, so we benefit from the deviations having non-trivial functional forms. If the deviations were totally independent of $\tau$, then the $Z^\prime$ would simply change the cross section by an overall factor. By the same reasoning, the closer the deviations are to a flat distribution in $\tau$, the the worse the constraint will be. One can see this in the case of the constraint due to only $\mu^+\mu^-$ in \cref{fig:finalStateLimits}, where the more inclusive $|\eta|$ cut of 2.5 counter-intuitively performs worse than the cut of 2.0, despite the more inclusive cut resulting in a greater total cross section. One can see in the upper-left panel of \cref{fig:dsdtall} that the deviation of $\partial_\tau\sigma$ from the SM is flatter as a function of $\tau$ for the cut of 2.5 than the cut of 2.0. This effect competes with the increased statistics from the larger acceptance. Still, for the limit using the combination of all final states, the marginalization finds a value of $\hat{\hat{\nu}}$ that leads to an improved constraint for the larger acceptance. The fact than increasing the acceptance can impact constraints in this non-trivial way indicates that we have lost information when we integrated over angular phase space. Thus, a more sophisticated analysis would benefit from probing the joint probability density in both $\tau$ and angle.

\section{Revisiting Direct Searches at 10 TeV}

The prospects for constraining an $L_{\mu} - L_{\tau}$ gauge boson at a high-energy muon collider were first evaluated in Ref.~\cite{Huang:2021nkl}. Ref.~\cite{Dasgupta:2023zrh} subsequently studied a broader set of channels that would be possible at such a machine.
These include constraints from 
$s$-channel $\mu^+\mu^- \to \mu^+\mu^-$ or $\tau^+\tau^-$, 
$Z'$-pair production with decays to muon or tau final states, 
or associated production of a $Z'$ and photon, with the $Z'$ subsequently decaying either to visible or invisible final states.
Of these, the $s$-channel production is most powerful for heavy $Z'$ masses, approaching the center of mass energy of the collider. 
For masses less than a few hundred GeV, the associated production channels are most constraining. 

In both Refs.~\cite{Huang:2021nkl} and \cite{Dasgupta:2023zrh}, only a $\sqrt{s} = 3~\textrm{TeV}$ muon collider was considered, and projections were only made for $M_{Z'} > 100~\textrm{GeV}$. 
In this section, we revisit the projections from these searches assuming $\sqrt{s} = 10~\textrm{TeV}$ with $10~\textrm{ab}^{-1}$ integrated luminosity for direct comparison with the indirect constraint we explored in the main text.
Associated production with the $Z'$ decaying to $\mu^+\mu^-$ suffers from the largest background at a muon collider, so we will focus on $\mu^+ \mu^- \to \tau^+\tau^- \gamma$ with a $\tau^+\tau^-$ resonance and $\mu^+\mu^- \to \nu\nu\gamma$---a ``missing mass'' search. 

\begin{itemize}

    \item Simulate in MadGraph, with cuts implemented at generator level, using an in-house model for the $L_{\mu} - L_{\tau}$ model. 

    \item Emphasize that we're okay with an aggressive estimate of the bounds; leads to conservative appraisal of the PDF approach

    \item checked that we reproduce the results of Ref.~\cite{Dasgupta:2023zrh} when using their cuts at $3~\textrm{TeV}$
    
\end{itemize}

\subsection{\texorpdfstring{$Z^\prime(\tau\tau) + \gamma$}{Z'(tautau) + gamma} Search}

First we consider associated $Z' \gamma$ production, with the $Z'$ decaying to $\tau^+\tau^-$ and appearing as a resonance. 
We require a single photon and two $\tau$ leptons, all with $p_T \geq 30~\textrm{GeV}$ and pseudorapidity $|\eta| < 2.44$. 
We further require all three final state particles to have an angular separation $\Delta R = \sqrt{\Delta\eta^2 + \Delta\phi^2} > 0.3$. 
With these cuts and isolation criteria, we follow Ref.~\cite{Dasgupta:2023zrh} and assume an overall $70\%$ efficiency for identifying the final state $\tau$ leptons and neglect any reconstruction effects on their momentum.
This includes both the hadronic and leptonic final states.
This should be considered an optimistic projection, as the $\tau$ decays will always include some invisible component from the neutrino, and their identification and reconstruction may be challenging. 

After these initial selection criteria, we compute the signal and Standard Model background---the latter including only the irreducible $\mu^+\mu^- \to \tau^+\tau^-\gamma$ process in the Standard Model---within an invariant mass window around the $Z'$ mass. For $M_{Z'} \geq 200~\textrm{GeV}$, we use a window of $\pm 0.05 \times M_{Z'}$, while for smaller masses we use a constant window of $\pm 10~\textrm{GeV}$.

This analysis, when performed with $\sqrt{s} = 3~\textrm{TeV}$, reproduces exactly the constraints shown in Fig.~15 of Ref.~\cite{Dasgupta:2023zrh}.  
We take this as a validation of the procedure and scale the center-of-mass energy to $10~\textrm{TeV}$ while also considering smaller masses. 
The resulting constraint is shown as an orange-dashed line in Fig.~\ref{fig:exclusion} of the main text. The bound is weaker between $\sim 75$ and $100~\textrm{GeV}$ due to the resonant SM $Z \to \tau^+\tau^-$ background. 
The constraint on $g'$ also gets gradually weaker for small masses, disappearing entirely below $\sim 50~\textrm{GeV}$, due to the requirement $\Delta R_{\tau\tau} > 0.3$. For small $Z'$ masses, the two decay products are nearly collinear, and cannot be resolved. A more detailed analysis, in which the $Z'$ is reconstructed as a single boosted object could extend the sensitivity to smaller masses. This would require a more detailed detector-level study, which we leave to future work.

\subsection{Monophoton Search}

We turn next to the monophoton channel, which is relevant for the $Z'$ decaying invisibly to neutrino final states. 
Here we require a single isolated photon with $p_{T,\gamma} \geq 30~\textrm{GeV}$ and $|\eta_{\gamma}| < 2.44$. 
The $Z'$ mass can be reconstructed using the ``recoil mass'' variable,
\begin{equation}
M^2_{\textrm{recoil}} = (p_1 + p_2 - p_{\gamma})^2 
= s - 2 \sqrt{s} E_{\gamma} \,,
\end{equation}
with $p_1$, $p_2$ the momenta of the initial state muons. 
We see that when the $Z'$ is taken on-shell the photon energy is monochromatic. 

In Ref.~\cite{Dasgupta:2023zrh}, a cut on the recoil mass $|M_{\textrm{recoil}} - M_{Z'}| < 10~\textrm{GeV}$ was imposed. 
For heavy $Z'$ masses and the lower center of mass energy considered there, this is a reasonable choice, and we reproduce their constraint.  
For the higher-energy collider and smaller $Z'$ masses we are interested in here, however, such a cut would require an extremely high resolution on photon energies, as at leading-order, the monochromatic signal is peaked very close to the beam energy of $5~\textrm{TeV}$. 
Therefore, we instead impose a cut on the photon energy directly, requiring it to be within $\pm 5~\textrm{GeV}$ of the value predicted from the recoil mass variable described above. For heavy $Z'$ masses, this results in essentially the same constraint as using the recoil mass.
Note that this projection is still quite optimistic: a resolution of 5~GeV is for the relevant photon energies is quite challenging. Moreover, this doesn't account for soft radiation from the incoming muons (parameterized by their PDFs), which would deplete the hard scattering energy and lead to a significant spread in the recoil mass. 

For $M_{Z'} \lesssim 250~\textrm{GeV}$, the Standard Model background with a resonant $Z \to \nu\nu$ is within the photon energy cut.
As a result, for all these masses the constraint on $g'$ is essentially flat, as the Standard Model background doesn't change as a function of the photon energy. This constraint becomes the leading bound for $M_{Z'} \lesssim 105~\textrm{GeV}$. 

\end{document}